\documentclass[final,5p,times,twocolumn]{elsarticle} 

\usepackage{hyperref}

\journal{Nuclear Instruments \& Methods in Physics Research, Section A}

\usepackage{ifthen} 
\newboolean{uprightparticles}
\setboolean{uprightparticles}{false} 
 \usepackage{xspace} 
 \usepackage{upgreek}

\def\lhcb {\mbox{LHCb}\xspace}

\def\cms    {\mbox{CMS}\xspace}

\ifthenelse{\boolean{uprightparticles}}%
{

 \def\PDelta      {\ensuremath{\Delta}\xspace}                 
 \def\PXi      {\ensuremath{\Xi}\xspace}                 
 \def\PLambda      {\ensuremath{\Lambda}\xspace}                 
 \def\PSigma      {\ensuremath{\Sigma}\xspace}                 
 \def\POmega      {\ensuremath{\Omega}\xspace}                 
 \def\PUpsilon      {\ensuremath{\Upsilon}\xspace}                 
 
 \def\PB      {\ensuremath{\mathrm{B}}\xspace}                 
                  
 \def\PD      {\ensuremath{\mathrm{D}}\xspace}

 \def\PK      {\ensuremath{\mathrm{K}}\xspace}

 \def\Pb      {\ensuremath{\mathrm{b}}\xspace}                 
 \def\Pc      {\ensuremath{\mathrm{c}}\xspace}

 \def\Pi      {\ensuremath{\mathrm{i}}\xspace}

}
{

 \mathchardef\PDelta="7101
 \mathchardef\PXi="7104
 \mathchardef\PLambda="7103
 \mathchardef\PSigma="7106
 \mathchardef\POmega="710A
 \mathchardef\PUpsilon="7107
                  
 \def\PB      {\ensuremath{B}\xspace}                 
                  
 \def\PD      {\ensuremath{D}\xspace}

 \def\PK      {\ensuremath{K}\xspace}

 \def\Pb      {\ensuremath{b}\xspace}                 
 \def\Pc      {\ensuremath{c}\xspace}

 \def\Pi      {\ensuremath{i}\xspace}

}

\def\cquark    {\ensuremath{\Pc}\xspace}

\def\ccbar     {\ensuremath{\cquark\cquarkbar}\xspace}
\def\bquark    {\ensuremath{\Pb}\xspace}

\def\bbbar     {\ensuremath{\bquark\bquarkbar}\xspace}

  \def\Kbar  {\kern 0.2em\overline{\kern -0.2em \PK}{}\xspace}

  \def\Dbar    {\kern 0.2em\overline{\kern -0.2em \PD}{}\xspace}

\def\Bbar    {\ensuremath{\kern 0.18em\overline{\kern -0.18em \PB}{}}\xspace}

  \def\Y#1S{\ensuremath{\PUpsilon{(#1S)}}\xspace}

\def\Lbar {\ensuremath{\kern 0.1em\overline{\kern -0.1em\PLambda}}\xspace}

\def\AT#1     {\ensuremath{A_{\mathrm{T}}^{#1}}\xspace}           

\def\C#1      {\ensuremath{\mathcal{C}_{#1}}\xspace}                       
\def\Cp#1     {\ensuremath{\mathcal{C}_{#1}^{'}}\xspace}                    
\def\Ceff#1   {\ensuremath{\mathcal{C}_{#1}^{\mathrm{(eff)}}}\xspace}        
\def\Cpeff#1  {\ensuremath{\mathcal{C}_{#1}^{'\mathrm{(eff)}}}\xspace}       
\def\Ope#1    {\ensuremath{\mathcal{O}_{#1}}\xspace}                       
\def\Opep#1   {\ensuremath{\mathcal{O}_{#1}^{'}}\xspace}                    

\newcommand{\tev}{\ifthenelse{\boolean{inbibliography}}{\ensuremath{~T\kern -0.05em eV}\xspace}{\ensuremath{\mathrm{\,Te\kern -0.1em V}}\xspace}}
\newcommand{\gev}{\ensuremath{\mathrm{\,Ge\kern -0.1em V}}\xspace}
\newcommand{\mev}{\ensuremath{\mathrm{\,Me\kern -0.1em V}}\xspace}
\newcommand{\kev}{\ensuremath{\mathrm{\,ke\kern -0.1em V}}\xspace}
\newcommand{\ev}{\ensuremath{\mathrm{\,e\kern -0.1em V}}\xspace}
\newcommand{\gevc}{\ensuremath{{\mathrm{\,Ge\kern -0.1em V\!/}c}}\xspace}
\newcommand{\mevc}{\ensuremath{{\mathrm{\,Me\kern -0.1em V\!/}c}}\xspace}
\newcommand{\gevcc}{\ensuremath{{\mathrm{\,Ge\kern -0.1em V\!/}c^2}}\xspace}
\newcommand{\gevgevcccc}{\ensuremath{{\mathrm{\,Ge\kern -0.1em V^2\!/}c^4}}\xspace}
\newcommand{\mevcc}{\ensuremath{{\mathrm{\,Me\kern -0.1em V\!/}c^2}}\xspace}

\def\mum  {\ensuremath{\,\upmu\rm m}\xspace}
\def\muma {\ensuremath{\,\upmu\rm m^2}\xspace}

\def\invfb   {\ensuremath{\mbox{\,fb}^{-1}}\xspace}

\def\degc {\ensuremath{^\circ}{C}\xspace}

\def\gsim{{~\raise.15em\hbox{$>$}\kern-.85em
          \lower.35em\hbox{$\sim$}~}\xspace}
\def\lsim{{~\raise.15em\hbox{$<$}\kern-.85em
          \lower.35em\hbox{$\sim$}~}\xspace}

\newcommand{\lum} {\ensuremath{\mathcal{L}}\xspace}

\def\nonn {\ensuremath{\rm {\it{n^+}}\mbox{-}on\mbox{-}{\it{n}}}\xspace}

\def\tell1  {TELL1\xspace}
\def\ukl1   {UKL1\xspace}

\def\cotwo         {\ensuremath{\rm CO_2}\xspace}

\begin{document}
\def\lhcb  {LHCb }
\def\bbbar {$\mathsf{b\overline{b}}$ }  
\def\ccbar {$\mathsf{c\overline{c}}$ }
\def\um {$\mathrm{\mu m}$ }
\def\umsq {$\mathrm{\mu m}^2$ }
\def\cotwo {$\mathrm{CO}_2$ }
\def\cms {$\mathrm{cm}^{-2}\mathrm{s}^{-1}$}
\def\copycern {\emph{\newline\copyright\  Copyright CERN: reproduced with permission}}
\def\copynikhef {\emph{\newline\copyright\  Copyright NIKHEF: reproduced with permission}}

\begin{frontmatter}

\title{LHCb VELO Upgrade}
\author{Karol Hennessy\fnref{myfootnote}}
\address{University of Liverpool, Liverpool, UK.}
\fntext[myfootnote]{on behalf of the LHCb VELO Upgrade collaboration}


\begin{abstract}
The upgrade of the LHCb experiment, scheduled for LHC Run-III, scheduled
to start in 2021, will transform the experiment to a trigger-less system
reading out the full detector at 40 MHz event rate. All data reduction
algorithms will be executed in a high-level software farm enabling the
detector to run at luminosities of $2\times10^{33}$\cms.

The Vertex Locator (VELO) is the silicon vertex detector surrounding
the interaction region. The current detector will be replaced with a
hybrid pixel system equipped with electronics capable of reading out
at 40 MHz. The upgraded VELO will provide fast pattern recognition and
track reconstruction to the software trigger. The silicon pixel sensors
have 55$\times$55\muma pitch, and are read out by the VeloPix ASIC,
from the Timepix/Medipix family. The hottest region will have pixel hit
rates of 900 Mhits/s yielding a total data rate of more than 3 Tbit/s
for the upgraded VELO. The detector modules are located in a separate
vacuum, separated from the beam vacuum by a thin custom made foil. The
foil will be manufactured through milling and possibly thinned further
by chemical etching.

The material budget will be minimised by the use of evaporative \cotwo
coolant circulating in microchannels within 400\mum thick silicon
substrates. The current status of the VELO upgrade is described and latest
results from operation of irradiated sensor assemblies are presented.
\end{abstract}


\end{frontmatter}


\section{LHCb and its Upgrade}
\begin{figure}[t]
        \begin{center}
                \includegraphics[width=0.45\textwidth]{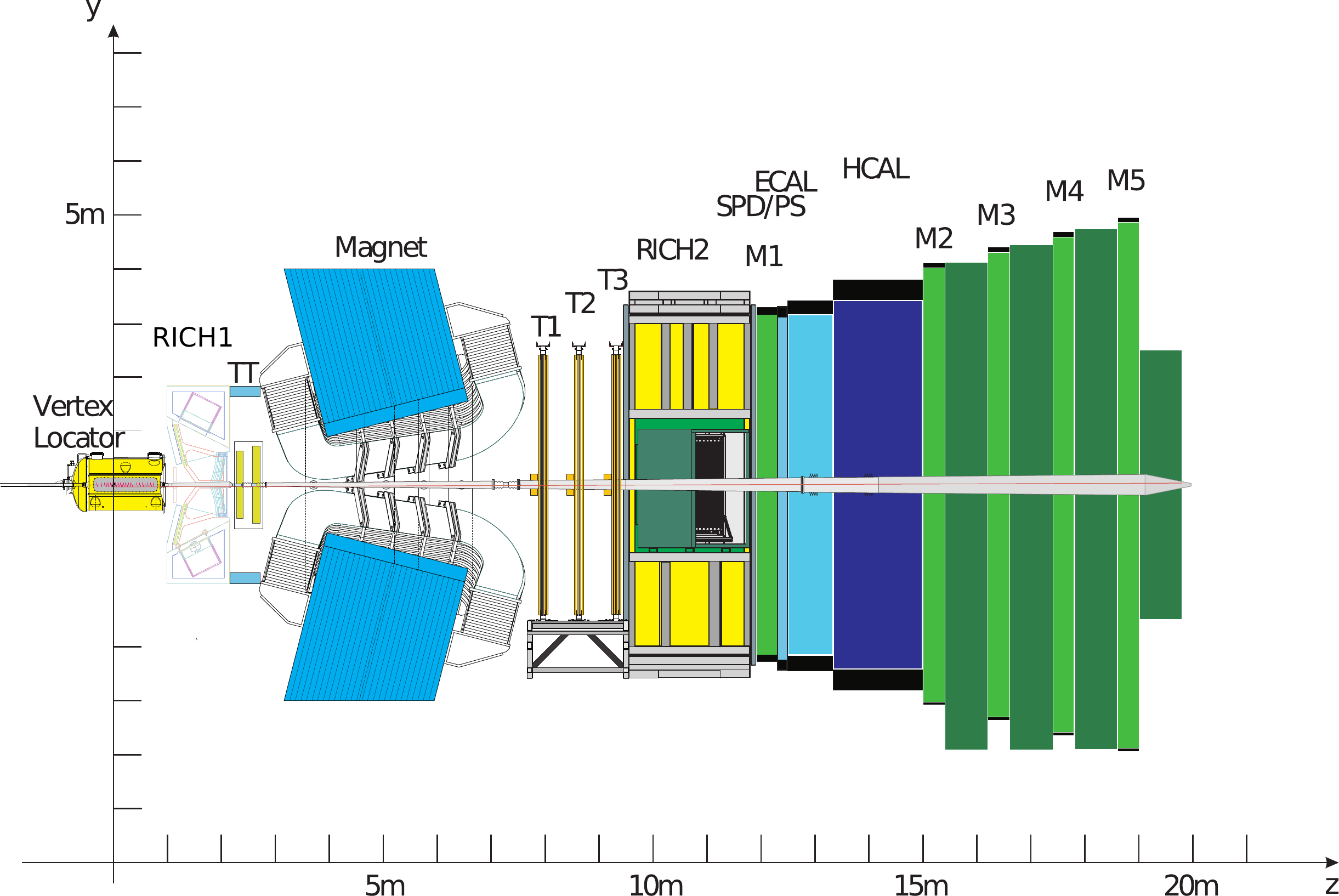} 
                \caption{
                        \label{fig:lhcb} 
                        The LHCb detector - starting at the interaction
                        point and proceeding downstream, we have the
                        Vertex Locator, RICH1, TT, Magnet, T1-T2-T3
                        tracking stations, RICH2, Calorimeters and
                        finally Muon chambers.
                        \copycern
                } 
        \end{center}
\end{figure}

The Large Hadron Collider Beauty detector \cite{lhcb} (figure
\ref{fig:lhcb}) is a flavour physics detector, designed to detect
decays of \bquark - and \cquark -hadrons for the study of CP violation and rare
decays.  In pp collisions at LHC \cite{lhc} energies, \bbbar
production is primarily in the forward/backward direction.  LHCb
has been designed as a forward arm spectrometer, to exploit this
fact.  LHCb is a precision experiment operating at a instantaneous 
luminosity of $\lum$ = 4$\times$10 $^{32}$\cms, and is
expected to continue in its current configuration until 2018.  

At the end of Run-II, many of the LHCb measurements will remain statistically 
dominated.
Luminosity will be increased by a factor of five
to $2\times 10^{33}$\cms in Run-III.  The increased luminosity at
LHCb will boosts statistics primarily in semi-leptonic channels containing at 
least one muon.  However, the trigger yield for purely hadronic
decay channels saturates due to energy cuts in the hardware trigger. 
Removing the hardware trigger eliminates the 1 MHz bottleneck and improves
the efficiency due to having full information available for the first
level trigger decision Running the detector at 40 MHz requires radical
changes to many of the subdetectors of LHCb.


\section{The Current VELO detector}
The VELO \cite{velo}\cite{kazu} is a silicon strip detector surrounding the interaction point at
LHCb.  The sensors are made of 300 \mum \nonn silicon, and are positioned only
7 mm from the beam line during data taking.  The VELO is split into two
halves such that each can be retracted during LHC injection.
The two VELO halves are operated in vacuum and separated from the primary LHC vacuum by means of a 300 \mum
thick aluminium foil.  
The VELO is required to have excellent
impact parameter resolution, essential for the reconstruction of heavy hadron decays.

The VELO consists of 42 modules placed along the beam direction, with
the full length of the detector being approximately 1m.  
A module is made of two half-disc sensors with R- and $\Phi$-measuring geometry. 

\section{VELO Upgrade}

\begin{figure}
        \begin{center} 
                \includegraphics[width=0.45\textwidth]{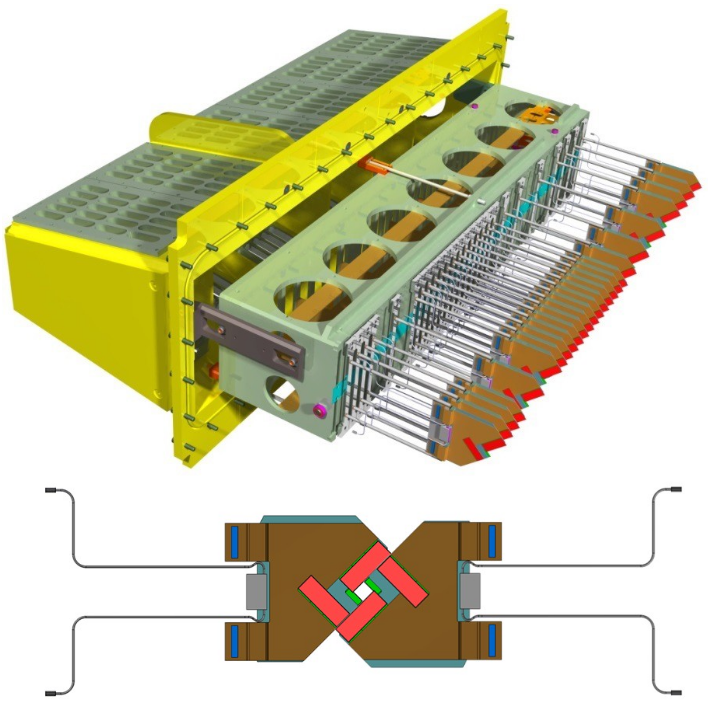}
                \caption{
                        \label{fig:velo-halves}
                        (top) Schematic of one half of the VELO upgrade detector.  Twenty-six modules 
                        are aligned along the beam direction.  (bottom) Two modules in the closed position.
                        The modules are in different z-positions along the beam line to minmise gaps in the 
                        acceptance and there is a slight overlap between the two halves which aids detector
                        alignment.
                        \copynikhef
                }
	\end{center}
\end{figure}
The Vertex Locator upgrade \cite{veloup} is a significant redesign from the original detector.  The 
major changes in new detector and its predecessor are as follows:
\begin{itemize}
        \item The detector will change from a silicon strip detector to a pixel detector.  
        \item The detector will be closer to the beam in its closed position at 5.1\,mm from 8.2\,mm
        \item The upgraded VELO will use a new VeloPix ASIC \cite{velopix} that can be read out at 
                40\,MHz (up from 1.1\,MHz) and a bandwidth of up to 20.4\,Gb/s.
        \item The current detector modules are cooled with \cotwo passing through 
                a series of cooling blocks attached to the base of the module substrate.  
                The upgraded modules will also use \cotwo as a coolant, but it will pass through micro-channels 
                in a  silicon substrate, directly beneath the major heat sources (VeloPix ASICs and other chips).
\end{itemize}

\begin{figure}
                \includegraphics[width=0.38\textwidth]{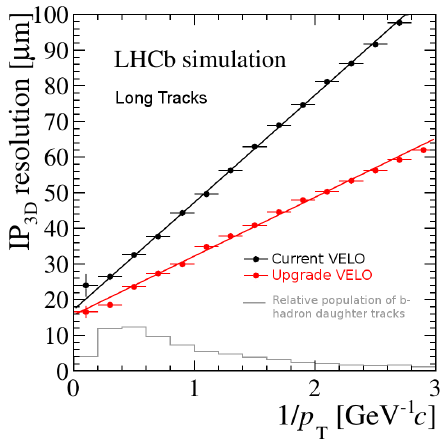} 
                \caption{
                        \label{fig:ip-resolution}
                        3D Impact Parameter resolution vs particle
                        inverse transverse momentum.  Made using LHCb
                        software simulation.  A significant resolution
                        improvement of the upgraded detector (red)
                        can be seen over the current VELO (black).
                        The upgrade operates closer to the beam line
                        and has a lower material budget - two major
                        contributions to the improved resolution.
                        \copycern
                }
\end{figure}

The upgraded VELO (shown in figure \ref{fig:velo-halves}) will have more
robust track reconstruction performance compared to a strip detector
and an overall improved resolution (figure \ref{fig:ip-resolution}).
This improved performance is achieved by (i) lowering the material budget with thinner
sensors and and thinner aluminium foil housing each detector half and
(ii) placing the sensors closer to the beam.  The closer placement of
the detector modules to the beam and the higher luminosity for LHC Run
III comes with some costs - the detector must be able to handle higher
radiation doses; there is a higher hit occupancy due to increased particle
flux.  This, in turn, results in higher data rates from the detector and
greater power consumption in the front-end ASICs (however, the dominant
increase in data rate is due to the removal of the L0 hardware trigger).
Table \ref{tab:upgrade-diffs}
lists some of the major differences between the current VELO detector
and the upgrade.

\begin{table}
        \begin{small}
        \begin{tabular}{l c c}
                \hline
                Feature & Current VELO & Upgraded VELO \\
                \hline
                Sensors & R \& $\phi$ strips & Pixels \\
                \# of modules & 42 & {52} \\
                Detector Active area & 0.22 m$^2$ & 0.12 m$^2$ \\
                        & $\sim$172k strips & $\sim$41M pixels\\
                Technology & electron collecting & electron collecting \\
                & 300\mum thick & {200\mum thick} \\
                Max fluence  & $3.9\times10^{14} $ & {$8\times10^{15} $} \\
                & $\mathrm{MeV\cdot n_{eq}/cm^{-2}}$ & {$\mathrm{MeV\cdot n_{eq}/cm^{-2}}$} \\
                HV tolerance & 500 V & {1000 V} \\
                ASIC Readout rate   & 1 MHz & 40 MHz \\
                Total data rate & $\sim$150 Gb/s & 1.2 Tb/s\\
                Total Power consumption & $\sim$ 1 kW & {2.2-2.3 kW} \\
                \hline
        \end{tabular}
        \end{small}
        \caption{An outline of the major differences between the current VELO and its upgrade}
        \label{tab:upgrade-diffs}
\end{table}

\subsection{Modules}
\begin{figure}[t]
        \centering
                \includegraphics[width=0.45\textwidth]{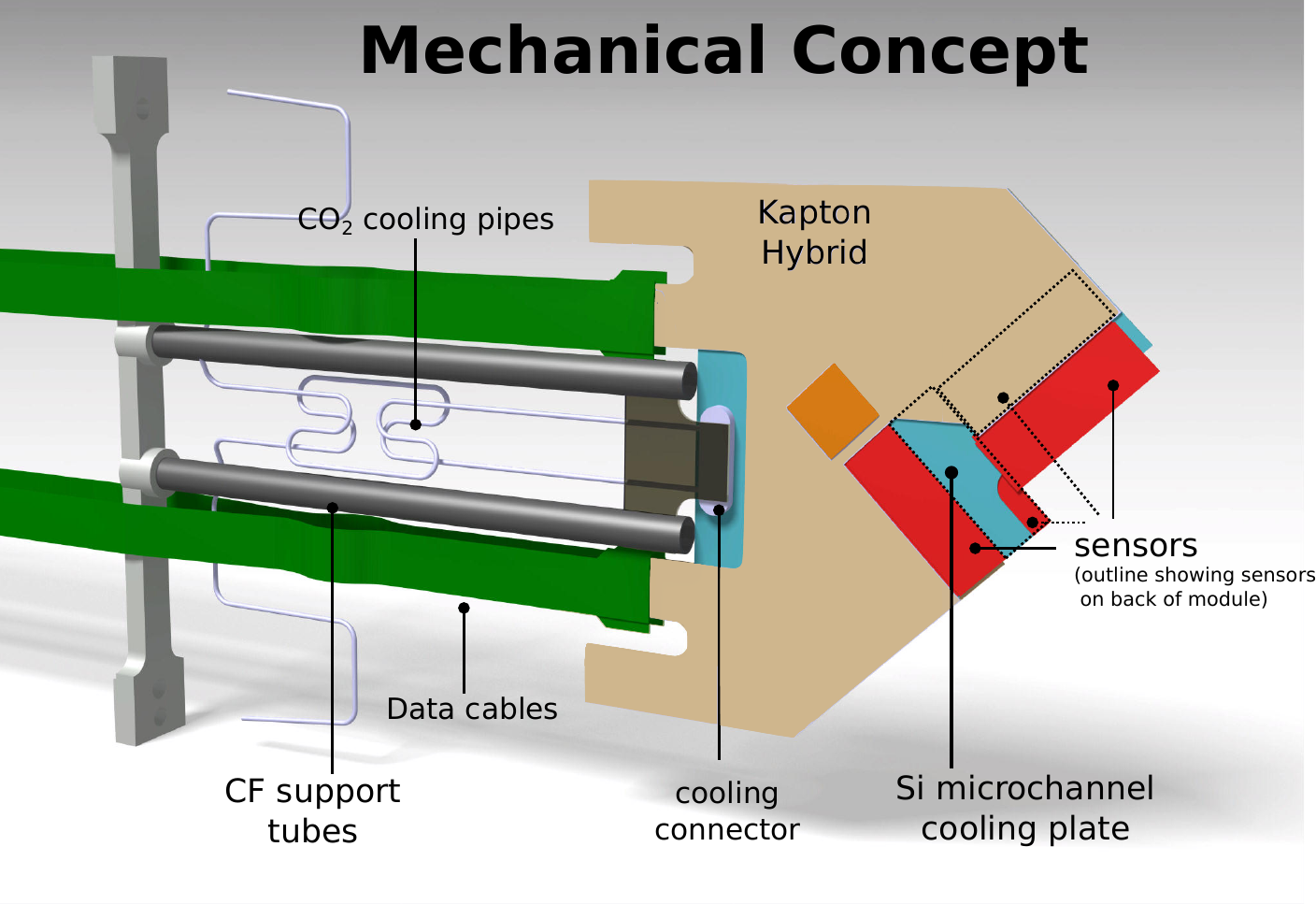}
                \caption{
                        A conceptual 3D model of the mechanical design of a  VELO upgrade module (subject to change).
                        \label{fig:velomodule}
                        \copynikhef
                }
\end{figure}

A mechanical module concept is shown in figure \ref{fig:velomodule}.
The current design of the module consists of a carbon-fibre structure
supporting a silicon microchannel substrate.  Stress-relieved \cotwo
cooling pipes route the \cotwo to and from the cooling connector which
is soldered to the Si substrate.  The module has four silicon sensors,
read out by, and bump-bonded to twelve VeloPix ASICs.  The ASICs are
glued directly onto the microchannel substrate.  Discrete electronics,
including a GBTX chip \cite{gbtx} for slow control, along with power, bias, and 
readout circuitry will be arranged on a Kapton hybrid.  
When the VELO is closed, the sensors
of opposing detector halves form a diamond shape with the beam passing
through the centre.

\subsection{Cooling}

The primary sources of heat on the module are due to the VeloPix chips.
It is estimated that the full chip will consume $<1.1$ W/cm$^{2}$.
The sensors must be kept at $-20$\degc in order to minimise the chance
of thermal runaway due to radiation damage.  The modules are cooled
using evaporative \cotwo at $-30$\degc.  The \cotwo is passed from 
an inlet pipe through a cooling connector soldered to the silicon
substrate.  The inlet fans out to a series of parallel microchannels
in the substrate which pass directly under the VeloPix chips.  Near the
inlet, the microchannels form a restriction region of higher pressure.
This is followed by a transition region where the channels widen from
60\mum$\times$ 60\mum to 120\mum$\times $200\mum stimulating boiling of
the \cotwo.  The microchannel layout has been optimised to route the
coolant directly to the site of the heat sources (the GBTX chip and
VeloPix ASICs) and minimise the temperature gradient across the module.
Several prototype substrates have been studied to evaluate their heat-load and
pressure performance.  The design was optimised to meet specifications in
terms of structural integrity and cooling performance (expected maximum cooling
power $>$36W).

\subsection{RF Foil}
\begin{figure}
        \includegraphics[width=0.45\textwidth]{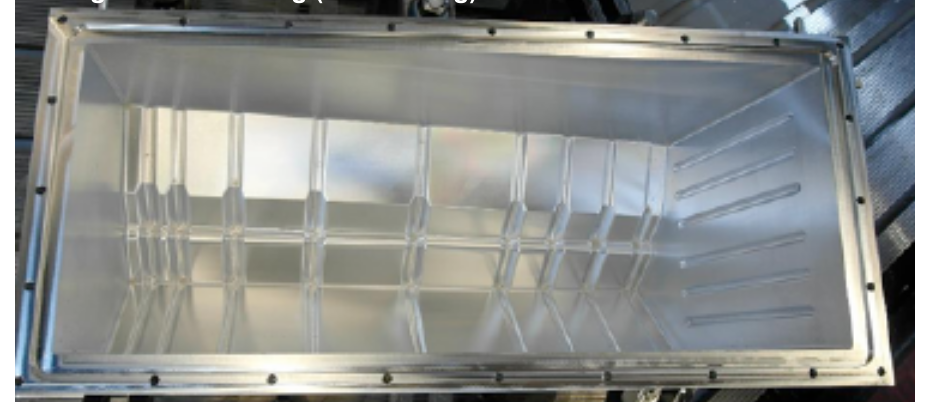}
        \caption{
                Half box prototype after milling.  The box starts as a
                180\,kg block of aluminium and ends as a 3\,kg piece.
                \label{fig:foil}
                \copynikhef
        }
\end{figure}

The aluminium foil serves to separate the secondary VELO vacuum from the
primary LHC vacuum.  The RF foil thickness has a significant impact
on the impact parameter resolution.  For the current detector, with a
300\mum foil, particles cross an average of $\sim3$\,mm of aluminium before the
first hit in the silicon sensor.  The proposed design for the upgrade
VELO foil will be 200\mum thick.  Studies on the foil manufacture have
been done - including milling the foil from a solid block 
of aluminium (a prototype half-box is show in figure \ref{fig:foil}), and
chemically etching areas of the foil which have greatest impact on the 
IP resolution.  

\section{Beam tests}

A campaign of beam tests were performed throughout 2014 and 2015
to characterise the performance of candidate VELO upgrade sensors.
The two major vendors studied were Hamamatsu (HPK) and
Micron.  Testing was performed with a Timepix3 telescope.  Timepix3 is
a precursor of the VeloPix ASIC but runs at a lower data rate (80\,Mhit/s,
compared to 900\,Mhits/s for VeloPix).
VeloPix has binary readout, whereas Timepix3 has analogue readout 
yielding a better time and spatial resolution.
Timepix3 has the same pixel geometry as VeloPix and therefore was an 
ideal test-bed for characterising the VELO upgrade sensors\footnote{The VeloPix
design had not been finalised at the time of sensor testing}.

\begin{figure}[t]
                \includegraphics[width=0.45\textwidth]{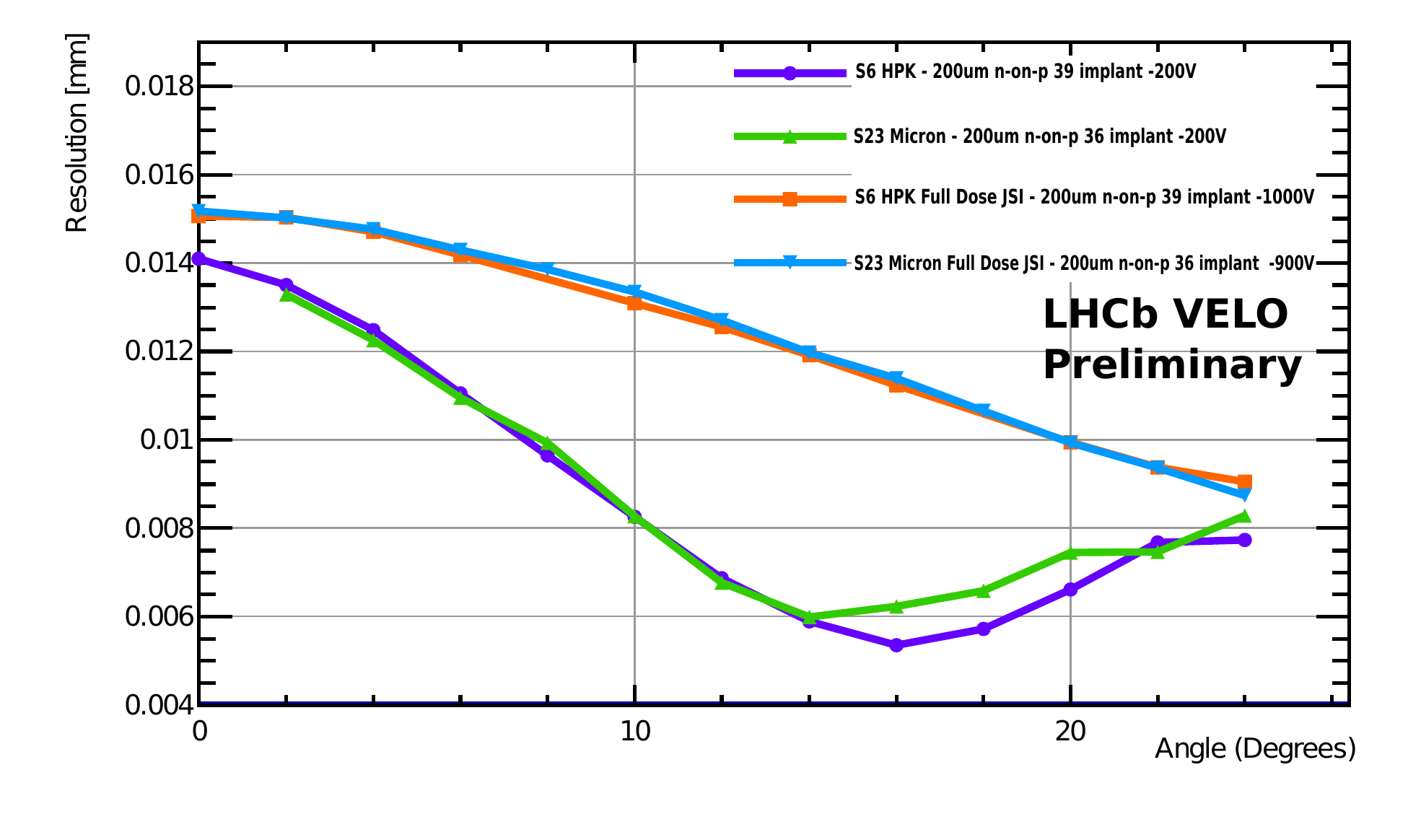}
                \caption{
                        Single hit resolution of two test sensors before (Micron - green, HPK - purple) and after
                        (Micron - blue, HPK - orange) irradiation as a function of incident beam angle.  Both
                        sensors show very similar performance.  An approximate factor of two degradation in 
                        resolution can be seen in the worst case.  The sensors were irradiated to $8\times10^{15} \mathrm{MeV\cdot n_{eq}/cm^{-2}}$.
                        \label{fig:resolution}
                        \copycern
                }
\end{figure}

\begin{figure}[t]
                \includegraphics[width=0.45\textwidth]{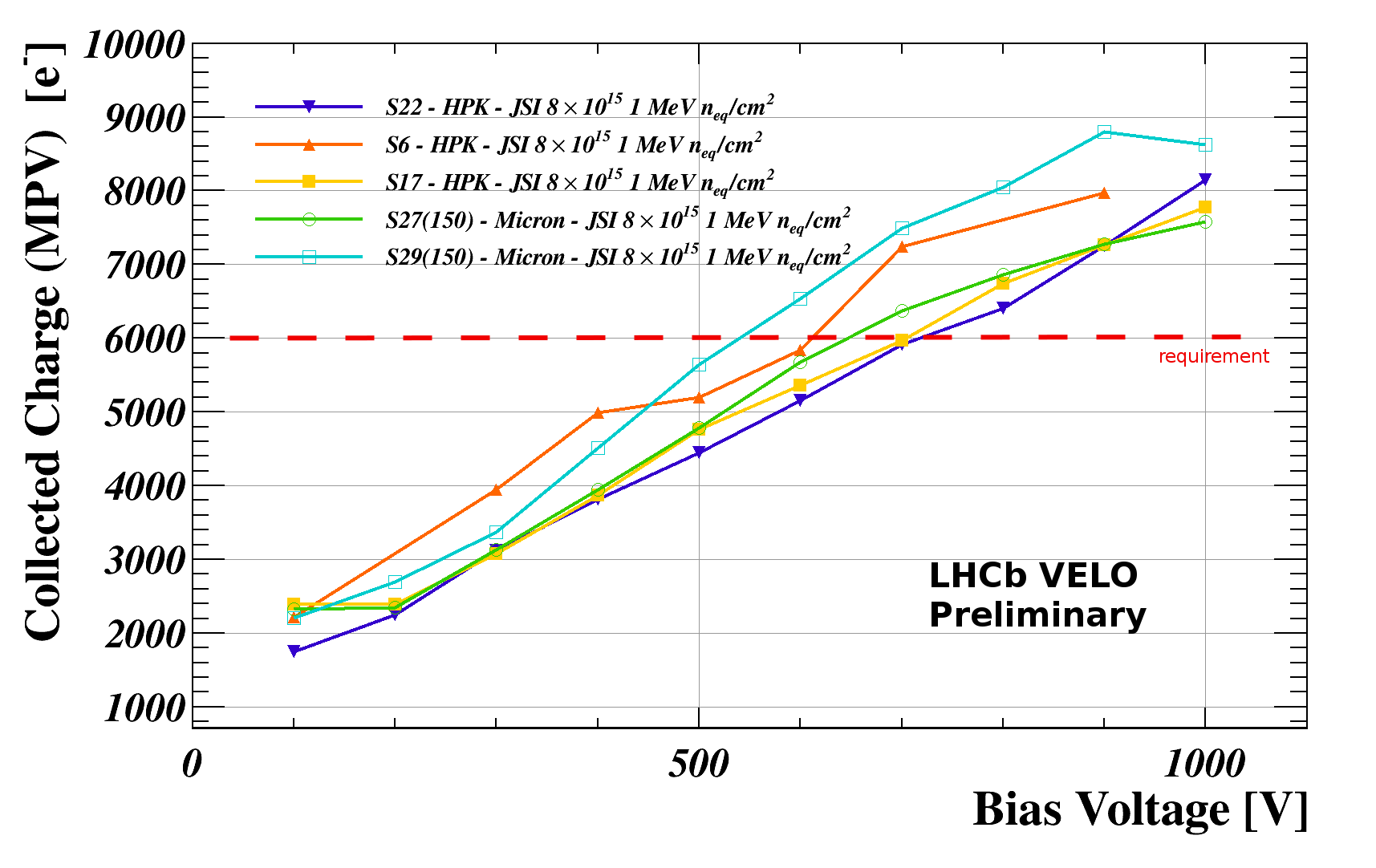}
                \caption{
                        Collected charge (determined from Most Probable Value of their Landau distributions) as
                        a function of Bias voltage after irradiation at full dose ($8\times10^{15}\, \mathrm{MeV\cdot n_{eq}/cm^{-2}}$).
                        Three Hamamatsu (200\mum) and two Micron (150\mum) sensors were tested.  All sensors pass the 6000 electron requirement 
                        (represented by the dashed line) at a voltage limit of 1000\,V.  
                        \label{fig:mpv}
                        \copycern
                }
\end{figure}

The upgrade sensors will receive an highly non-uniform radiation dose up to
$8\times10^{15}\,\mathrm{MeV\cdot n_{eq}/cm^{-2}}$ during their lifetime.
At this dose, the sensors are expected to retain a 99\% hit efficiency
at up to 1000\,V bias voltage without suffering breakdown.  The HPK and Micron
sensors tested with the Timepix3 telescope were compared un-irradiated
and irradiated to their maximum dose ($8\times10^{15}\,\mathrm{MeV\cdot n_{eq}/cm^{-2}}$ 
equivalent to an integrated luminosity of 50\invfb).  The
single hit resolution of detectors is shown in figure \ref{fig:resolution}
and the charge collection performance is shown in figure \ref{fig:mpv}.
Sensor tests defined a baseline choice of 200\mum n-on-p silicon as
suitable for the VELO upgrade.  This is subject to change if, for
example, a similarly performing thinner sensor be found.

\section{VeloPix and DAQ}

At the heart of new VELO upgrade electronic design is the new VeloPix
ASIC \cite{velopix}.  The VeloPix is based on the Timepix3 ASIC and has
a data-driven readout.  The $256\times256$ pixel array is arranged into groups of $2\times 4$ pixels
called SuperPixels.  Binary hit information is time-stamped, addressed and read out as SuperPixels packets.  Arranging the data in this way reduces the output bandwidth by $\approx 30\%$ over 
standard pixel readout.  The data readout latency varies as a function of SuperPixel hit
position, and hits closer to the beam take longer to reach the end-of-column 
readout logic.  In addition, the hits will be disordered in time.

The VeloPix will be read out using a PCIe40 readout board \cite{pcie40}
which has been designed as a generic readout board for the LHCb upgrade.
The PCIe40 firmware is designed as a series of common components with the
option for user-specific data processing.  The common components deal
with accepting the input data from the detector over the GBT protocol
\cite{gbt}, error-checking, dealing with reset signals, and preparing the
data for the event builder and computing farm.   The user-specific code
would, for example, perform zero-suppression or similar data-reduction
techniques.

For the VELO upgrade, some of the common input blocks have had to be
replaced as the GBT protocol is not used for reasons of power consumption
at the front-end.  A simpler protocol - Gigabit Wireline Transmitter
(GWT) has been used instead.  The VELO user-specific blocks include primarily
(i) data descrambling, decoding and parity checking,
(ii) a time-ordering router which re-sequences the data coming from the
continuous readout of the VeloPix chip, and (iii) a hit isolation tagger
which performs a simplified type of clusterisation to reduce the load
downstream in the reconstruction software running on the CPU farm.
Resource utilisation within the FPGA is within acceptable tolerances
and studies of packet loss rates are ongoing.  Preparations are being
made to assess VeloPix readiness.

\section{Conclusion}

The VELO upgrade detector makes significant improvements on its
predecessor.  Improved tracking performance is achieved using a pixel
sensor closer to the interaction region (5.1\,mm) but with a reduced
material budget.  This leads to an even higher and non-uniform radiation
tolerance requirement.  As such the sensors and readout chips are
cooled using a \cotwo coolant via a microchannel silicon substrate.
Candidate sensors meeting the required performance constraints have been
identified. Significant progress has been made to both control and read out
the forthcoming VeloPix chip which will be tested in the coming months.

\section*{Acknowledgements}
Thanks to Science Technology Funding Council (STFC) of the UK, University of Liverpool, CERN and associated institutes of LHCb.

\bibliography{mybibfile}

\end{document}